\documentstyle[12pt,amssym]{article}
\def\C{\mbox{$\Bbb C$}}
\def\R{\mbox{$\Bbb R$}}

\def\case#1#2{{\textstyle{#1\over #2}}}
\def\sech{\mathop{\rm sech}\nolimits}

\title{
Pseudo-Hermiticity, weak pseudo-Hermiticity and $\eta$-orthogonality condition}

\author{B.\ Bagchi $^{a,}$\thanks{E-mail: bbagchi@cucc.ernet.in}\ , C.\ Quesne
$^{b,}$\thanks{Directeur de recherches FNRS; E-mail: cquesne@ulb.ac.be} \\
{\small \sl $^a$ Department of Applied Mathematics, University of Calcutta,} \\
{\small \sl 92 Acharya Prafulla Chandra Road, Calcutta 700 009, India}\\
{\small \sl $^b$ Physique Nucl\'eaire Th\'eorique et Physique
Math\'ematique,  Universit\'e Libre de Bruxelles,} \\ {\small \sl Campus de la
Plaine CP229, Boulevard~du Triomphe, B-1050 Brussels, Belgium}}
\date{ }
\begin{document}
\baselineskip=22pt plus 1pt minus 1pt
\maketitle

\begin{abstract}
We discuss certain features of pseudo-Hermiticity and weak pseudo-Hermiticity
conditions and point out that, contrary to a recent claim, there is no inconsistency if the
correct orthogonality condition is used for the class of pseudo-Hermitian, PT-symmetric
Hamiltonians of the type $H_{\beta} = [p + {\rm i} \beta \nu(x)]^2/2m + V(x)$.
\end{abstract}

\vspace{0.5cm}


 
\bigskip\noindent
Corresponding author: C.\ Quesne, Physique Nucl\'eaire Th\'eorique et Physique
Math\'e\-ma\-ti\-que,  Universit\'e Libre de Bruxelles, Campus de la Plaine
CP229, Boulevard du Triomphe, B-1050 Brussels, Belgium

\noindent
Telephone: 32-2-6505559

\noindent
Fax: 32-2-6505045

\noindent
E-mail: cquesne@ulb.ac.be 
\newpage
%
%
In recent times it has been stressed that neither Hermiticity nor PT symmetry serves as a
necessary condition for a quantum Hamiltonian to preserve the reality of its bound-state
eigenvalues~\cite{bessis, bender98, bender99, dorey, mosta02a}. In fact, it has been
realized~\cite{mosta02a} that the existence of real eigenvalues can be associated with a
non-Hermitian Hamiltonian provided it is $\eta$-pseudo-Hermitian:
\begin{equation}
  \eta H = H^{\dagger} \eta,  \label{eq:pseudo}
\end{equation}
where $\eta$ is a Hermitian linear automorphism and, assuming $\hbar = 2m = 1$,
\begin{equation}
  H = p^2 + V(x)  \label{eq:H}
\end{equation}
for $V(x) \in \C$ and $p = - {\rm i} \partial_x$. Then, in such a case, the spectrum of a
diagonalizable $H$ is real if there exists a linear invertible operator $O$ such that $\eta =
(O O^{\dagger})^{-1}$. Moreover, one can relax $H$ to be only weak
pseudo-Hermitian~\cite{solombrino} by not restricting $\eta$ to be Hermitian.\par
%
%
The purpose of this Letter is to establish the following results:

\noindent (i) The twin concepts of pseudo-Hermiticity and weak pseudo-Hermiticity are
complementary to one another.

\noindent (ii) For a first-order differential realization, $\eta$ may be anti-Hermitian but for
the second-order case, $\eta$ is necessarily Hermitian. For both cases, we make
connections to the same PT-symmetric Scarf II Hamiltonian (having normalizable
eigenfunctions) to show that the choice of $\eta$ is not unique in ascertaining the
character of the Hamiltonian.

\noindent (iii) For the class of $\eta$-pseudo-Hermitian, PT-symmetric Hamiltonians
described by~\cite{mosta02b, ahmed02}~\footnote{Note that in Ref.~\cite{ahmed02},
it is assumed that $\hbar = m = 1$ instead of $\hbar = 2m = 1$.}
\begin{equation}
  H_{\beta} = [p + {\rm i} \beta \nu(x)]^2 + V(x), \qquad \beta \in \R,  \label{eq:Hbeta}
\end{equation}
where the odd function $\nu(x) \in \R$, $V(x)$ is PT-symmetric, and 
\begin{equation}
  \eta = \exp\left[- 2\beta \int^x \nu(y) dy\right],  \label{eq:eta}
\end{equation}
our earlier derivation~\cite{bagchi01} of the generalized continuity equation for
Hamiltonians of the form~(\ref{eq:H}) [with $V(x)$ PT-symmetric] can be extended to
$H_{\beta}$ as well. The resulting $\eta$-orthogonality condition needs to be
implemented judiciously.\par
%
%
We begin by addressing to the point (i) above. Consider some non-Hermitian $\eta$ that
is subject to the condition~(\ref{eq:pseudo}). Taking Hermitian conjugate, we obtain, on
adding and subtracting, the following combinations
\begin{equation}
  \eta_+ H = H^{\dagger} \eta_+, \qquad \eta_- H = H^{\dagger} \eta_-, 
  \label{eq:eta+-} 
\end{equation}
where $\eta_{\pm} = \eta \pm \eta^{\dagger}$. While the first one of~(\ref{eq:eta+-})
corresponds to strict pseudo-Hermiticity, the second one points to weak
pseudo-Hermiticity with a new anti-Hermitian operator $\eta_-$. Note that $\eta_+$ is
Hermitian. It is thus clear that weak pseudo-Hermiticity is not more general than
pseudo-Hermiticity but works complementary to it.\par
%
%
We now turn to (ii). Decomposing $V(x)$ and $\eta$ as
\begin{eqnarray}
  V(x) & = & V_R(x) + {\rm i} V_I(x), \nonumber \\
  \eta & = & \frac{d}{dx} + f(x) + {\rm i} g(x),  \label{eq:order1}
\end{eqnarray}
where $V_R$, $V_I$, $f$, $g \in \R$, we get, on inserting~(\ref{eq:order1})
into~(\ref{eq:pseudo}), the relations
\begin{eqnarray}
  V_I & = & {\rm i} (f' + {\rm i}g'), \nonumber \\
  V'_R + {\rm i} V'_I & = & - (f'' + {\rm i}g'') - 2{\rm i} V_I (f + {\rm i}g). 
          \label{eq:order1-inter}
\end{eqnarray} 
In (\ref{eq:order1-inter}) the primes denote the order of differentiations with respect to
the variable $x$. We are then led to the conditions
\begin{equation}
  V'_R = - 2gg', \qquad cg' = 0, \qquad c \in \R,
\end{equation}
which imply the existence of two solutions corresponding to $c=0$ and $g'=0$,
respectively. In the following we concentrate on the case $c=0$ because $g'=0$ yields a
trivial result that corresponds to a real constant potential with no normalizable
eigenfunction.\par
%
%
{}For the choice $c=0$, it turns out that
\begin{equation}
  f = 0, \qquad V_R = - g^2 + k, \qquad V_I = - g',
\end{equation}
where $k \in \R$. In consequence, we have the results
\begin{eqnarray}
  V(x) & = & - g^2(x) + k - {\rm i} g'(x), \nonumber \\
  \eta & = & \frac{d}{dx} + {\rm i} g(x).  \label{eq:order1bis}
\end{eqnarray}
The above form of $V(x)$ shows that, in the framework of supersymmetric quantum
mechanics, we can associate to it an imaginary superpotential $W(x) = {\rm i} g(x)$, its
partner being the complex conjugate potential. We also observe that, for even $g$
functions, the representation of $\eta$ makes it anti-Hermitian in character. Let us
consider the following specific example for $g = d \sech x$, $d \in \R$. We get
from~(\ref{eq:order1bis})
\begin{eqnarray}
  V(x) & = & - d^2 \sech^2 x + k + {\rm i} d \sech x \tanh x, \nonumber \\
  \eta & = & \frac{d}{dx} + {\rm i} d \sech x.  \label{eq:Scarf1}
\end{eqnarray}
\par
%
%
It is obvious that $V(x)$ is a particular case of the generalized PT-symmetric Scarf~II
potential investigated previously by us~\cite{bagchi00a} in connection with the complex
algebra sl(2,~\C). A comparison with the results obtained there shows that, in the present
case, we have a single series of real eigenvalues with normalizable eigenfunctions
provided $d > \frac{1}{2}$. The corresponding Hamiltonian is both P-pseudo-Hermitian and
$\eta$-weak-pseudo-Hermitian with $\eta$ given by~(\ref{eq:Scarf1}). Our example
confirms the assertion~\cite{mosta02c} that, for a given non-Hermitian Hamiltonian,
there could be infinitely many $\eta$ satisfying the weak-pseudo-Hermiticity or the
pseudo-Hermiticity condition.\par
%
%
We next attend to a second-order differential representation of $\eta$:
\begin{equation}
  \eta = \frac{d^2}{dx^2} - 2 p(x) \frac{d}{dx} + b(x),  \label{eq:order2}
\end{equation}
where $p$, $b \in \C$. Substituting (\ref{eq:order2}) into the condition
(\ref{eq:pseudo}), we obtain the constraints
\begin{eqnarray}
  b & = & - p' + p^2 - \frac{p''}{2p} + \left(\frac{p'}{2p}\right)^2 +
         \frac{\gamma}{4p^2}, \nonumber \\
  V & = & 2p' + p^2 + \frac{p''}{2p} - \left(\frac{p'}{2p}\right)^2 -
         \frac{\gamma}{4p^2} - \delta, \nonumber \\
  V^* & = & - 2p' + p^2 + \frac{p''}{2p} - \left(\frac{p'}{2p}\right)^2 -
         \frac{\gamma}{4p^2} - \delta,  \label{eq:order2-inter}
\end{eqnarray}
where $\gamma$, $\delta \in \R$. From the last two relations in (\ref{eq:order2-inter}),
it is clear that $p(x)$ must be pure imaginary,
\begin{equation}
  p(x) = {\rm i} a(x),
\end{equation}
where $a(x) \in \R$. As such $V(x)$ and $\eta$ acquire the forms
\begin{eqnarray}
  V(x) & = & 2{\rm i} a' - a^2 + \frac{a''}{2a} - \left(\frac{a'}{2a}\right)^2 +
         \frac{\gamma}{4a^2} - \delta, \nonumber \\
  \eta & = & \frac{d^2}{dx^2} - 2{\rm i} a(x) \frac{d}{dx} + b(x),  \label{eq:order2bis}  
\end{eqnarray}
with $b(x) = - V(x) + {\rm i} a' - 2 a^2 - \delta$. In (\ref{eq:order2bis}), $\eta$ can
be easily recognized to be a Hermitian operator since it can be written in the form $\eta
= - \tilde{O}^{\dagger} \tilde{O}$, where $\tilde{O} = \frac{d}{dx} + r - {\rm i} a$,
$\tilde{O}^{\dagger} = - \frac{d}{dx} + r + {\rm i} a$, and $r^2 - r' = \frac{a''}{2a} -
\left(\frac{a'}{2a}\right)^2 + \frac{\gamma}{4a^2}$. In Ref.~\cite{fityo}, such a
decomposition of $\eta$ was assumed, a priori, to arrive at some non-Hermitian
Hamiltonians with real spectra.\par
%
%
Let us, however, confine ourselves to the following choice
\begin{equation}
  a(x) = - \case{1}{2} B (2A+1) \sech x, \qquad \gamma = 0, \qquad \delta =
  \case{1}{4},
\end{equation}
where $A + \frac{1}{2} > 0$, $B > 0$, and $A - B + \frac{1}{2}$ is not an integer. We
are again led to the PT-symmetric Scarf~II potential having a more general form than
obtained with the first-order differential realization of $\eta$:
\begin{equation}
  V(x) = - V_1 \sech^2 x - {\rm i} V_2 \sech x \tanh x,  \label{eq:Scarf2} 
\end{equation}
where $V_1 = \frac{1}{4}[B^2 (2A+1)^2 + 3] > 0$ and $V_2 = - B (2A+1) \ne 0$.
According to Refs.~\cite{ahmed01, bagchi02}, the condition for real eigenvalues for the
Hamiltonian corresponding to (\ref{eq:Scarf2}) is $|V_2| \le V_1 + \frac{1}{4}$. Here it
amounts to $[B (2A+1) - 2]^2 \ge 0$, which is always met.\par
%
%
Of particular interest is the special case $B=1$:
\begin{equation}
  V(x) = - (A^2+A+1) \sech^2 x + {\rm i} (2A+1) \sech x \tanh x.  \label{eq:Scarf2-BR}
\end{equation}
On setting $A + \frac{1}{2} = - \lambda$ ($\lambda < 0$), Eq.~(\ref{eq:Scarf2-BR})
can be seen to reduce to the potential $V^{(1)} - \frac{1}{4}$ of Ref.~\cite{bagchi00b}
for $\mu=1$. The associated energy levels of (\ref{eq:Scarf2-BR})
are~\cite{bagchi00a}: $E^{(-\lambda)}_n = - (\lambda + n + \frac{1}{2})^2$ and
coincide with $E^{(2)}_n - \frac{1}{4}$ of~\cite{bagchi00b}. Note that there is, in
general, a doubling of energy levels in transiting from the real to the PT-symmetric
Scarf~II potential. In fact, the second algebra of sl(2, \C) leads to an additional energy
level $E^{(1)}_0 = - \frac{1}{4}$ that is consistent with the zero-energy state
of~\cite{bagchi00b}.\par
%
%
{}Finally, we take up a general derivation of the continuity equation for the class of
Hamiltonians $H_{\beta}$ given by (\ref{eq:Hbeta}). The associated Schr\"odinger
 equation reads
\begin{equation}
  {\rm i} \frac{\partial \psi(x,t)}{\partial t} = - \left(- \frac{\partial}{\partial x} + \beta
  \nu(x)\right)^2 \psi(x,t) + V(x) \psi(x,t).  \label{eq:SE1}
\end{equation}
From this it follows that the function $\psi^*(-x,t)$ satisfies
\begin{equation}
  - {\rm i} \frac{\partial \psi^*(-x,t)}{\partial t} = - \left(- \frac{\partial}{\partial x} + 
  \beta \nu(x)\right)^2 \psi^*(-x,t) + V(x) \psi^*(-x,t).  \label{eq:SE2}
\end{equation}
\par
%
%
On considering Eq.~(\ref{eq:SE1}) for some solution $\psi_1(x,t)$ and
Eq.~(\ref{eq:SE2}) for some other solution $\psi_2(x,t)$ and then multiplying
(\ref{eq:SE1}) and (\ref{eq:SE2}) by $\exp\left[- 2\beta \int^x \nu(y) dy\right]
\psi^*_2(-x,t)$ and $\exp\left[- 2\beta \int^x \nu(y) dy\right] \psi_1(x,t)$,
respectively, we obtain, on subtracting, a natural generalization of the continuity equation
for PT-symmetric quantum mechanics to its $\eta$-pseudo-Hermitian extension, namely
\begin{equation}
  \frac{\partial P_{\eta}(x,t)}{\partial t} + \frac{\partial J_{\eta}(x,t)}{\partial x} = 0, 
  \label{eq:CE}
\end{equation}
where
\begin{eqnarray}
  P_{\eta}(x,t) & = & \eta \psi^*_2(-x,t) \psi_1(x,t), \nonumber \\
  J_{\eta}(x,t) & = & \frac{\eta}{{\rm i}} \left[\psi^*_2(-x,t) \frac{\partial
         \psi_1(x,t)}{dx} - \psi_1(x,t) \frac{\partial \psi^*_2(-x,t)}{dx}\right],
\end{eqnarray}
and $\eta$ is defined in (\ref{eq:eta}). If $\psi_1(x,t) \to 0$ and $\psi_2(x,t) \to 0$ as
$x \to \pm \infty$, as is normally expected for bound-state wave functions, then
integration of (\ref{eq:CE}) over the entire real line gives the conservation law
\begin{equation}
  \frac{\partial}{\partial t} \int_{-\infty}^{\infty} dx\,  \eta\, \psi^*_2(-x,t) \psi_1(x,t) =
  0.  \label{eq:conservation} 
\end{equation}
\par
%
%
In the case of energy eigenfunctions
\begin{equation}
  \psi_1(x,t) = u_1(x) e^{-{\rm i} E_1 t}, \qquad \psi_2(x,t) = u_2(x) e^{-{\rm i} E_2 t},
\end{equation}
corresponding to the eigenvalues $E_1$ and $E_2$ respectively,
Eq.~(\ref{eq:conservation}) reduces to
\begin{equation}
  (E_1 - E_2^*) \int_{-\infty}^{\infty} dx\, \eta\, u_2^*(-x) u_1(x) = 0.
  \label{eq:eta-ortho}
\end{equation}
Equation (\ref{eq:eta-ortho}) represents the $\eta$-orthogonality
condition~\cite{mosta02a}. Obviously it transforms to the
PT-orthogonality~\cite{bagchi01, japaridze}
\begin{equation}
  (E_1 - E_2^*) \int_{-\infty}^{\infty} dx\, u_2^*(-x) u_1(x) = 0
  \label{eq:PT-ortho}
\end{equation}
for $\nu(x) = 0$. Indeed Eq.~(\ref{eq:eta-ortho}) can be derived from (\ref{eq:PT-ortho})
by effecting a gauge transformation on the wave functions $u$ of $H$ in a manner $u
\to \exp[- \int^x \beta \nu(y) dy] u$, $V(x)$ being PT-symmetric. As such $H_{\beta}$
may be looked upon as a gauge-transformed version of $H$. However, it needs to be
emphasized that care should be taken to correctly implement the normalization
conditions deriving from (\ref{eq:eta-ortho}) and (\ref{eq:PT-ortho}) and which are
appropriate to the Hamiltonians $H_{\beta}$ and $H$, respectively. Thus although
PT-symmetric, the form of the $\eta$-pseudo-Hermitian Hamiltonian $H_{\beta}$ at once
suggests that the normalization condition related to (\ref{eq:eta-ortho}) is to be used
rather than that connected with (\ref{eq:PT-ortho}), a point overlooked in
Ref.~\cite{ahmed02}.\par
%
%
In summary, we have shown that $\eta$-pseudo-Hermiticity and weak pseudo-Hermiticity
are essentially complementary concepts. We have provided an explicit example of
PT-symmetric Scarf~II model to demonstrate that $\eta$ does not necessarily have a
unique representation to determine the character of the associated non-Hermitian
Hamiltonian. We have also pointed out the correct use of the $\eta$-orthogonality
condition when dealing with a pseudo-Hermitian gauge-transformed Hamiltonian.\par
%
%

\end{document}